\documentstyle[11pt,newpasp,twoside,epsf]{article}
\def\arg#1{{\it#1\/}}

\def\edcomment#1{\iffalse\marginpar{\raggedright\sl#1\/}\else\relax\fi}
\reversemarginpar

\begin{document}
\title{The Kinematics of Point-Symmetric Planetary Nebulae:
Observational
Evidence of Precessing Outflows}
\author{Mart\'{\i}n A. Guerrero}
\affil{Astronomy Department, University of Illinois, 1002 W. Green
Street, Urbana, IL 61801, USA}

\begin{abstract}
The discovery of collimated outflows associated to point-symmetric
features in Planetary Nebulae has proliferated in recent years. 
The systematic variation of radial velocity that many of them show
strongly suggests a uniform rotation or precession of the ejection 
direction. 
Although several physical processes have been invoked, the formation
mechanism of precessing collimated outflows in PNe is currently an 
intriguing but unresolved problem. 
\end{abstract}

\vspace*{-0.5cm}
\section{The Point-Symmetry Morphology}

Point-symmetry in Planetary Nebulae (PNe) was introduced as a main 
morphological class by Corradi, Schwarz, \& Stanghellini (1993) as: 
``those objects [$\dots$] whose morphology shows no other symmetries
than 
point-reflection about the central source''. 
The increasing discovery of point-symmetric components in PNe has made
clear 
that point-symmetry in PNe span nearly all the morphological classes
and 
evolutionary stages. 
Among the many different point-symmetric features now recognized to be 
common in PNe, it can be quoted:

\noindent
$\bullet$ The {\it ansae} or FLIERs (Fast Low-Ionization Emission Regions),  
          which appear to be outflows departing from the tips of the
major 
          axis of elliptical PNe (e.g. Balick et al.\ 1998). 

\noindent
$\bullet$ The straight elongated jets or string of knots, as in Hb\,4
(L\'opez, 
          Steffen, \& Meaburn 1997b) and NGC\,3918 (Corradi et al.\
2000). 
          
\noindent
$\bullet$ The marked point-symmetric brightness of the bipolar lobes
of 
          certain bipolar PNe, as is the case of Hb\,5 (e.g. Riera
1999).  
          
\noindent
$\bullet$ The multiple bipolar outflows at different position angles
and with 
          different degrees of collimation reported in the quadrupolar
PNe 
          M\,2-46 and M\,1-75 (Manchado, Stanghellini, \& Guerrero
1996) or 
          in the poly-polar PN NGC\,2440 (Pascoli 1987; L\'opez et
al.\ 1998). 
          
\noindent
$\bullet$ The multiple point-symmetric pairs of components found at
different 
          position angles and distances to the center of the PN. 
          Such is the case of the collimated outflows in NGC\,6543
(Miranda 
          \& Solf 1992; Harrington \& Borkowski 1995), and Fleming\,1,
the 
          prototype of BRET (Bipolar Rotating Episodic jeT) (L\'opez,
Meaburn, 
          \& Palmer 1993), that shows a string of knots bent in
opposite 
          directions on both sides of the central nebula. 

The peculiar spatial distribution of the pairs of knots in these PNe
is 
specially interesting, as it strongly suggests rotation in the
direction 
of the ejection.
I will focus on these PNe.

\section{Kinematics of Point-Symmetric PNe: Point-Symmetric Outflows}

At the same time that the sample of PNe with point-symmetric
components 
has been growing, kinematical information on a significant number of 
these PNe has been obtained. 
Information on the kinematics of point-symmetric morphological
components 
of PNe is very valuable, as it provides us with information of the
motions 
along the line of sight. 
This can be combined with the 2-D spatial distribution to construct a 
spatio-kinematic model.

\begin{table}
\caption{\arg{The Kinematics of Point-Symmetric PNe}}
\begin{tabular}{lccc}
\noalign{\smallskip}
\tableline
\noalign{\smallskip}
\multicolumn{1}{l}{Object\hspace*{1.0cm}} & 
\multicolumn{1}{c}{\hspace*{1.0cm}$v_{\rm r}$\hspace*{1.0cm}}   & 
\multicolumn{1}{c}{\hspace*{1.0cm}FWHM\hspace*{1.0cm}}          & 
\multicolumn{1}{c}{\hspace*{1.0cm}References\hspace*{1.0cm}} \\
\noalign{\smallskip}
\multicolumn{1}{l}{}       & 
\multicolumn{1}{c}{[km~s$^{-1}$]} & \multicolumn{1}{c}{[km~s$^{-1}$]}
& 
\multicolumn{1}{c}{}       \\
\noalign{\smallskip}
\tableline
\noalign{\smallskip}
NGC\,6543  & ~~25--~~40 &     10     &  1 \\
IC\,4634   & ~~30--~~35 &   $\dots$  &  2 \\
           &    100     &   $\dots$  &  3 \\	       
                                       
He\,2-186  &    135     &   $\dots$  &  2 \\
Fleming\,1 & ~~~7--~~75 & ~~17--~~26 &  4 \\
He\,3-1475 &  425--870  &  150--400  &  5,6,7 \\
Hu\,2-1    & ~~~2--~~56 & ~~16--~~29 &  8 \\
NGC\,6210  & ~~~5--~~21 &  $\dots$   &  9 \\
IC\,4593   & ~~~3--100  & ~~12--~~20 & 10,11 \\
KjPn\,8    &$\sim$0--220&  100--280  & 12 \\
MyCn\,18   &  200--500  & ~~20--~~40 & 13 \\
NGC\,6881  & ~~~1--~~~9 & ~~13--~~19 & 14 \\
He\,1-1    & ~~~9--~~43 &  $\dots$   & 15 \\
PC\,19     & ~~30--~~35 &  $\dots$   & 15 \\
Pe\,1-17   & ~~~2--~~24 &  $\dots$   & 15 \\
NGC\,6884  & ~~14--~~38 &     18     & 16 \\
NGC\,6572  & ~~~7--~~38 &     17     & 17 \\ 
K\,1-2     & ~~10--~~20 & ~~11--~~20 & 18 \\
Wray\,17-1 & ~~15--~~70 &  $\dots$   & 18 \\
K\,3-35    &     20     &     27     & 19 \\
M\,1-16    &   250      &  230       & 20 \\
M\,2-46    & ~~25--~~40 &            & 21 \\
NGC\,2440  &  100--150  &            & 22 \\
\tableline
\tableline
\noalign{\smallskip}
\noalign{\smallskip}
\end{tabular}
(1) Miranda \& Solf 1992; 
(2) Schwarz 1992b; 
(3) Hajian et al.\ 1997; 
(4) L\'opez, Meaburn, \& Palmer 1993; 
(5) Riera et al.\ (1995); 
(6) Bobrowsky et al.\ 1995; 
(7) Harrington 1999; 
(8) Miranda 1995; 
(9) Phillips \& Cuesta 1996; 
(10) Corradi et al.\ 1997; 
(11) O'Connor et al.\ 1999; 
(12) L\'opez et al.\ 1997a; 
(13) Bryce et al.\ 1997; 
(14) Guerrero \& Manchado 1998; 
(15) Guerrero, V\'azquez, \& L\'opez 1999; 
(16) Miranda, Guerrero, \& Torrelles 1999; 
(17) Miranda et al.\ 1999; 
(18) Corradi et al.\ 2000; 
(19) Miranda et al.\ 2000; 
(20) Schwarz 1992a; 
(21) Manchado, Stanghellini, \& Guerrero 1996; 
(22) L\'opez et al.\ 1998
\end{table}

Table~1 compiles the list of point-symmetric PNe with multiple pairs
of 
components for which valuable kinematical information is available. 
The quadrupolar and poly-polar PNe M\,1-16, M\,2-46, and NGC\,2440
have 
been included. 
The measured radial velocities and line widths are given in columns 
2 and 3 respectively.

In most of the cases, the radial velocities in Table~1 are in the
range 
$0-50$ km~s$^{-1}$. 
These are the typical velocity shifts reported in FLIERs (e.g. Balick
et 
al.\ 1994) and in straight elongated jets. 
Real expansion velocities must be larger because projection effects. 
On the other hand, KjPn\,8, MyCn\,18, and He\,3-1475 show quite
noticeable 
large radial velocities that may correspond to deprojected expansion 
velocities larger than 1\,000 km~s$^{-1}$.

The line width is typically $10-20$ km~s$^{-1}$. 
The narrow line width, together with the high expansion velocity and
small
spatial extension, prove that these structures are collimated
outflows. 
There are only two cases (KjPn\,8 and He\,3-1475) in which the width
is 
larger than 100 km~s$^{-1}$ revealing important dynamical effects.

Finally, one of the most important result from all these observations
is 
that the spatial distribution of morphological components is related
to 
systematic variations of their radial velocities in most of the cases. 
The term ``bipolar outflow'', commonly used in PNe, can therefore be 
generalized to ``point-symmetric outflow'' in these PNe.

\section{Precession in Planetary Nebulae}

The coupled variations of the spatial distribution and radial
velocities 
of pairs of point-symmetric knots strongly suggest a systematic
rotation 
of the ejection direction. 
Precession, the uniform change in orientation of an axis around a
fixed 
axis, is a very appealing phenomenon, as it involves very peculiar
physical 
scenarios that will be described in $\S4$. 
In the past, precession was invoked in order to explain the helical 
structures reported in NGC\,6543 (M\"unch 1968) and NGC\,7293, the 
Helix Nebula (Fabian \& Hansen 1979), and the overlapping bipolar 
structures of the poly-polar PN NGC\,2440 (Kaler \& Aller 1974;
Pascoli 
1987). 
More recently, different numerical simulations (Raga, Cant\'o, \& Biro 
1993; Raga \& Biro 1993; Cliffe et al.\ 1995; Cliffe, Frank, \& Jones 
1996) have shown that precessing jets can reproduce the morphological 
structure of point-symmetric PNe.


\begin{figure}
\plotfiddle{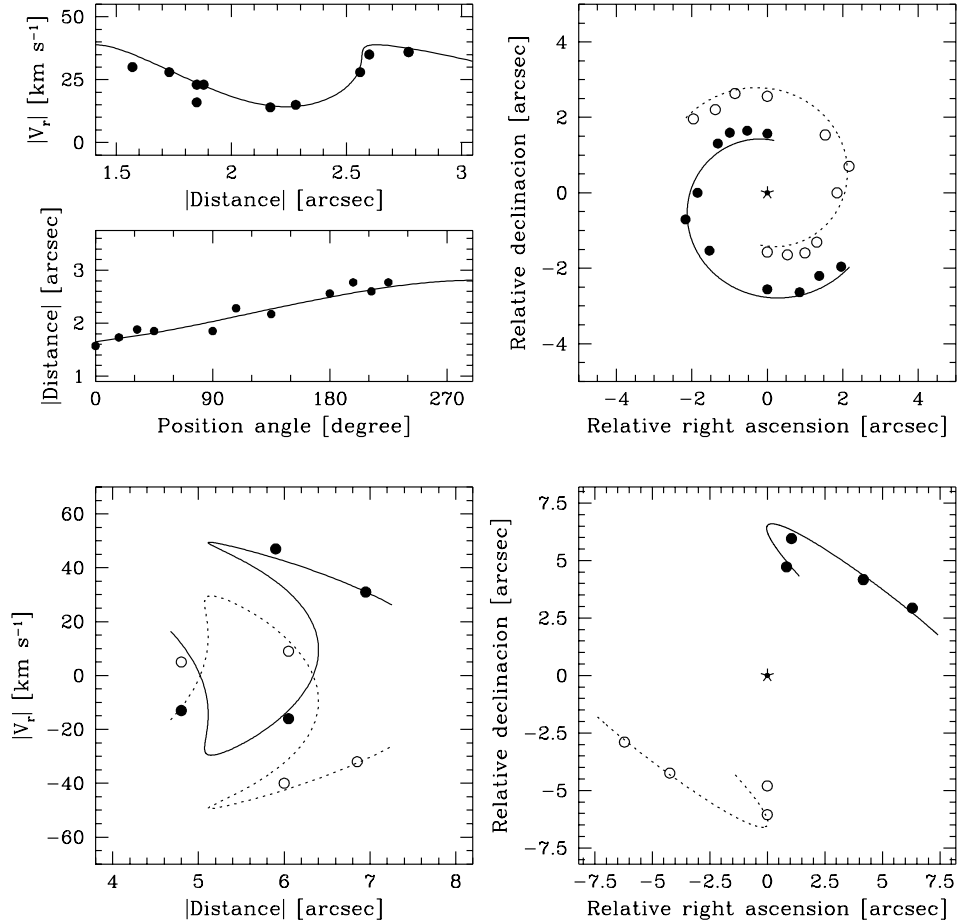}{9.0cm}{0}{65}{65}{-210}{-115}
\caption{[{\it top}] Relative variation of the radial velocity,
distance and 
position angle [{\it left}], and spatial distribution [{\it right}] of
the 
point-symmetric knots of NGC\,6884. 
[{\it bottom}] Relative variation of the radial velocity {\it versus} 
distance [{\it left}], and spatial distribution [{\it right}] of the 
knots of He\,1-1. 
In both cases, the lines represent the results of a ballistic
precessing 
jet model (see Table~2 for further details on the specific precession 
parameters and expansion velocity). 
}
\end{figure}

In a precessing collimated outflow, the variations of the radial
velocity 
and spatial position are determined by the following parameters: \\
$\bullet$ The aperture angle of the precession cone, $\Phi$. \\
$\bullet$ The inclination of the precession axis with the line of
sigth, 
          $i$. \\
$\bullet$ The precession period, $T$, or rotation rate, $d\omega/dt$.
\\
$\bullet$ The expansion velocity, $v_{\rm exp}$. \\
$\bullet$ The distance to the PN, $d$. \\
$\bullet$ The initial position angle of the ejection, $\omega_{\rm
o}$. \\
Depending on these parameters, the morphology and variation of radial 
velocity may be very different. 
For small inclination angles with the line of sight, a double helix or 
spiral is expected. 
This is the case of NGC\,6884 (Miranda et al.\ 1999). 
Variations of the radial velocities are small, but change
systematically 
with the position angle or distance (Figure~1-{\it top}). 
If the inclination angle increases, so the radial velocity differences
do. 
Such is the case of He\,1-1 (Figure~1-{\it bottom}), with a large 
inclination angle, but a smaller aperture angle than NGC\,6884
(Guerrero 
et al.\ 1999). 
A large inclination and small aperture angle produce a loop-like
structure, 
as is observed in NGC\,6881 (Guerrero \& Manchado 1998) or an S-shaped 
structure, as in He\,3-1475 (Borkowsky, Blondin, \& Harrington 1997).

It is interesting to note that the steady increase of the radial 
velocity with distance has been reported for Fleming\,1 (L\'opez 
et al.\ 1993) and MyCn\,18 (Bryce et al.\ 1997). 
A ballistic precessing outflow can reproduce this behavior, but only
for large inclination and aperture angles,\footnote{ 
This completely justify the simplification of a BRET model in which
the ejection direction rotation is limited into a plane.} 
and very restrictive initial ejection direction.  
An alternative explanation has recently been proposed by
Garc\'{\i}a-Segura et al.\ (1999) and will be discussed in more 
detail in $\S$4.

Using the uniform precessing ballistic model above outlined, the
precession 
parameters ($\Phi$, $v_{\rm exp}$, $i$, and $T$) can be worked out by 
fitting the observational data. 
Figure~1 shows the fit to the data of NGC\,6884 and He\,1-1. 
The available precession parameters are summarized in Table~2 where
$\Delta t$ 
stands for the fraction of the period, $T$, that the ejection lasted. 
Although these estimates are subject to uncertainties in the distance, 
assumptions in the model, and resolution of the observations and model 
fitting to them, the wide range of parameters is real. 
Interestingly, the ejection lasts only a fraction of the period 
in all the cases.

\begin{table}
\caption{\arg{Precession Parameters of Point-Symmetric PNe}}
\begin{tabular}{lccccl}
\noalign{\smallskip}
\tableline
\noalign{\smallskip}
\multicolumn{1}{l}{Object} & 
\multicolumn{1}{c}{$2\times\Phi$} & 
\multicolumn{1}{c}{$v_{\rm exp}$} & 
\multicolumn{1}{c}{$T$} & 
\multicolumn{1}{c}{$\Delta t$} & 
\multicolumn{1}{l}{Remarks} \\
\noalign{\smallskip}
\multicolumn{1}{l}{} & 
\multicolumn{1}{c}{[degree]} & 
\multicolumn{1}{c}{[km~s$^{-1}$]} & 
\multicolumn{1}{c}{[yr]} & 
\multicolumn{1}{c}{[yr]} & 
\multicolumn{1}{l}{} \\
\tableline
\noalign{\smallskip}  
NGC\,6881  &   44   &   11    & $3\,800\times d$ & 0.8 & \\
NGC\,6884  &  120   &   55    &      500         & 0.6 & \\
Fleming\,1 &  180   &   85    &    195\,000      & 0.08& BRET model \\
           &  130   &  110    &    100\,000      & 0.14& \\
MyCn\,18   &  180   & $>1000$ &      4\,700      & 0.11& BRET model \\
           &  160   &  1000   &      9\,500      & 0.07& \\
He\,1-1    &   70   &    75   &   $250\times d$  & 0.4 & \\ 
NGC\,6543  & 26--70 & 40--200 &   500--25\,000   & 0.25-0.45 & \\
M\,2-46    & $>53$  & 25--40  & $>18\,500$       & $\dots$ & \\
\tableline
\tableline
\end{tabular}
\end{table}

There are still many point-symmetric PNe with kinematical information
for 
which the available data do not allow to perform an accurate
determination 
of the precession parameters. 
These are the cases of those PNe in which only two pairs of
point-symmetric 
components are detected (for example, NGC\,6543). 
In such cases, the parameter space can be restricted using additional 
constraints provided by a detailed morphological description. 
Therefore, the combination of ground-based high-dispersion
spectroscopic 
observations with high spatial resolution HST WFPC2 narrow-band images 
can make a dramatic improvement in the spatio-kinematic modeling.

\section{Physical Scenarios}

Several models have been proposed to explain the formation of
precessing 
jets in PNe. 
They can be grouped into three different categories:

\vspace*{0.1cm}
\noindent$\bullet$ Hydrodynamical focusing + wobbling instabilities.
\\ 
      It has been proposed that weakly collimated bipolar outflows in
      PNe are focused into jets by oblique radiative shocks (Frank,
      Balick, \& Livio 1996). 
      Borkowski et al.\ (1997) used a similar mechanism to explain the 
      formation of the symmetric pairs of knots observed in He\,3-1475 
      invoking additional hydrodynamical instabilities and/or asymmetries 
      of the confining medium. 
      While this mechanism might work for He\,3-1475, it seems
difficult to 
      apply to precession motions characterized by large aperture
angles. 
      In addition, it seems quite unlikely that such instabilities may 
      reproduce the point-symmetric distribution of pairs of knots. 
      
\vspace*{0.1cm}
\noindent$\bullet$ Magnetic collimation around a precessing star. \\
      The magnetic tension may also be able to produce collimated
flows or 
      jets (R\'o\.zyczka \& Franco 1996). 
      Using this scenario, Garc\'{\i}a-Segura (1997) explains the
formation 
      of point-symmetric structures by adding a rotating star whose
rotation 
      axis is misaligned to the magnetic field. 
      Although the resulting morphology (and presumably the
kinematics) in 
      this model results disturbed by the confining medium, the motion
can 
      still be interpreted as a precessing jet if the density of the
ejection 
      is larger than that of the confining medium. 
      More recently, Garc\'{\i}a-Segura et al.\ (1999) have shown that 
      magnetic collimation can naturally reproduce the high velocity
and 
      steady increase of radial velocity observed in MyCn\,18 and
Fleming\,1. 
      
\vspace*{0.1cm}
\noindent$\bullet$ Precessing or wobbling accretion disks. \\
      Livio \& Pringle (1996; 1997) have suggested that instabilities
in an 
      accretion disk during a common envelope phase may cause it to
wobble. 
      The short fraction of the period for which precession is
observed in PNe 
      does not allow to reject this mechanism. 
      Similarly, precession of the inner portion of an accretion disk
has also 
      been proposed by Morris \& Reipurth (1990) to explain the
point-symmetry 
      observed in IRAS\,09371+1212, and by Manchado et al.\ (1996) to
explain 
      the formation of quadrupolar PNe.

\section{Summary and Conclusions}

\begin{enumerate}
\item The kinematics of point-symmetric pairs of components in PNe
      exhibit systematic variations of the observed radial velocities 
      coupled with the spatial distribution. 
      These are compatible with a uniformly precessing episodic
      ballistic jet in most of the cases. 
\item The properties of the ejection show a wide range of values: \\
      \centerline {500 yr $< T < 2\times10^5$ yr} \\ 
      \centerline {$22^\circ < \Phi < 80^\circ$~~~~~} \\
      \centerline {11 km~s$^{-1} < v_{\rm exp} < 1000$ km~s$^{-1}$~~}
\item The ejection of material lasts for a fraction ($0.1-0.8$) of 
      the precession period. 
\end{enumerate}

At this moment, there is no definite answer to the question of what
mechanism 
(probably involving binary stars or magnetic fields) produces
point-symmetric 
outflows in PNe, but it is clear that precession or uniform rotation
of the 
ejection direction is present in these PNe. 
Further observations (combining high-dispersion spectroscopic
observations 
and high-resolution narrow-band images) are required to perform a
detailed 
characterization of point-symmetric outflows in PNe. 
The spatio-kinematic models of a larger sample of such outflows would
help 
to constraint the precession properties and to restrict the formation 
processes.

\acknowledgments
I thank G. Garc\'{\i}a-Segura, J. A. L\'opez, and L. F. Miranda for 
useful comments and discussion, and A. Riera for making available to 
me her spectral observations of He\,3-1475. 
This research is partially supported by the Direci\'on General de
Ense\~nanza 
Superior e Investigaci\'on Cient\'{\i}fica of the Spanish Ministerio
de 
Educaci\'on y Cultura.

\end{document}